\documentstyle[epsf,rotate,german]{mn}
\newif\ifAMStwofonts

\def\O{$\Omega$}
\input wasyfont

\def\ar{{$\rightarrow$}}

\def\D{{$\Delta$ }}

\ifoldfss
  \ifCUPmtlplainloaded \else
    \NewTextAlphabet{textbfit} {cmbxti10} {}
    \NewTextAlphabet{textbfss} {cmssbx10} {}
    \NewMathAlphabet{mathbfit} {cmbxti10} {} 
    \NewMathAlphabet{mathbfss} {cmssbx10} {} 
  \fi
  \ifAMStwofonts
    \ifCUPmtlplainloaded \else
      \NewSymbolFont{upmath} {eurm10}
      \NewSymbolFont{AMSa} {msam10}
      \NewMathSymbol{\upi}     {0}{upmath}{19}
      \NewMathSymbol{\umu}     {0}{upmath}{16}
      \NewMathSymbol{\upartial}{0}{upmath}{40}
      \NewMathSymbol{\leqslant}{3}{AMSa}{36}
      \NewMathSymbol{\geqslant}{3}{AMSa}{3E}

    \fi
  \fi
\fi 

\ifnfssone
  \newmathalphabet{\mathit}
  \addtoversion{normal}{\mathit}{cmr}{m}{it}
  \addtoversion{bold}{\mathit}{cmr}{bx}{it}
  \newmathalphabet{\mathbfit} 
  \addtoversion{normal}{\mathbfit}{cmr}{bx}{it}
  \addtoversion{bold}{\mathbfit}{cmr}{bx}{it}
  \newmathalphabet{\mathbfss} 
  \addtoversion{normal}{\mathbfss}{cmss}{bx}{n}
  \addtoversion{bold}{\mathbfss}{cmss}{bx}{n}
  \ifAMStwofonts
    \ifCUPmtlplainloaded \else
      \UseAMStwoboldmath
      \makeatletter
      \new@mathgroup\upmath@group
      \define@mathgroup\mv@normal\upmath@group{eur}{m}{n}
      \define@mathgroup\mv@bold\upmath@group{eur}{b}{n}
      \edef\UPM{\hexnumber\upmath@group}
      \new@mathgroup\amsa@group
      \define@mathgroup\mv@normal\amsa@group{msa}{m}{n}
      \define@mathgroup\mv@bold\amsa@group{msa}{m}{n}
      \edef\AMSa{\hexnumber\amsa@group}
      \makeatother
      \mathchardef\upi="0\UPM19
      \mathchardef\umu="0\UPM16
      \mathchardef\upartial="0\UPM40
      \mathchardef\leqslant="3\AMSa36
      \mathchardef\geqslant="3\AMSa3E
    \fi
  \fi
\fi 

\ifnfsstwo
  \DeclareMathAlphabet{\mathbfit}{OT1}{cmr}{bx}{it}  
  \SetMathAlphabet\mathbfit{bold}{OT1}{cmr}{bx}{it}
  \DeclareMathAlphabet{\mathbfss}{OT1}{cmss}{bx}{n}
  \SetMathAlphabet\mathbfss{bold}{OT1}{cmss}{bx}{n}
  \ifAMStwofonts
    \ifCUPmtlplainloaded \else
      \DeclareSymbolFont{UPM}{U}{eur}{m}{n}
      \SetSymbolFont{UPM}{bold}{U}{eur}{b}{n}
      \DeclareSymbolFont{AMSa}{U}{msa}{m}{n}
      \DeclareMathSymbol{\upi}{0}{UPM}{"19}
      \DeclareMathSymbol{\umu}{0}{UPM}{"16}
      \DeclareMathSymbol{\upartial}{0}{UPM}{"40}
      \DeclareMathSymbol{\leqslant}{3}{AMSa}{"36}
      \DeclareMathSymbol{\geqslant}{3}{AMSa}{"3E}
    \fi
  \fi
\fi 

\ifCUPmtlplainloaded \else
  \ifAMStwofonts \else 
    \def\upi{\pi}
    \def\umu{\mu}
    \def\upartial{\partial}
  \fi
\fi

\title{Orbital dynamics of three-dimensional bars: \\IV. 
       Boxy isophotes in face-on views}
\author[P.A.~Patsis et al.]
{P.A.~Patsis,$^1$ Ch.~Skokos,$^{1,2}$ E.~Athanassoula$^3$\\
$^1$Research Center of Astronomy, Academy of Athens, Anagnostopoulou 14,
  GR-10673 Athens, Greece\\
$^2$Division of Applied Analysis, Department of Mathematics and Center for
Research and Application of Nonlinear Systems (CRANS),\\ University of Patras,
GR-26500 Patras, Greece\\
$^3$Observatoire de Marseille, 2 Place Le Verrier, F-13248 Marseille Cedex 4,
  France}
\date{Accepted .
      Received ;
      in original form }
\pagerange{\pageref{firstpage}--\pageref{lastpage}}
\pubyear{2001}

\begin{document}

\maketitle

\label{firstpage}

\begin{abstract}
  We study the conditions that favour boxiness of isodensities in the
  face-on views of orbital 3D models for barred galaxies. Using
  orbital weighted profiles we show that
  boxiness is in general a composite effect that appears when one considers
  stable orbits belonging to several families of periodic orbits. 3D
  orbits that are introduced due to vertical instabilities, play a
  crucial role in the face-on 
  profiles and enhance their rectangularity. This happens because at the 4:1
  radial resonance region we have several orbits with boxy face-on
  projections, instead of few rectangular-like x1 orbits, which, in
  a fair fraction of the models studied so far,  
  are unstable at this region. Massive bars are characterized by
  rectangular-like orbits. However, we find that it is the pattern speed  
  that affects most the elongation of the boxy feature, in
  the sense that
  fast bars are more elongated than slow ones. Boxiness in intermediate
  distances between the center of the model and the end of the bar can be
  attributed to x1v1 orbits, or to a combination of families related to the
  radial 3:1 resonance.
\end{abstract}


\section{Introduction}    
\begin{figure}
\hspace{1cm} \epsfbox{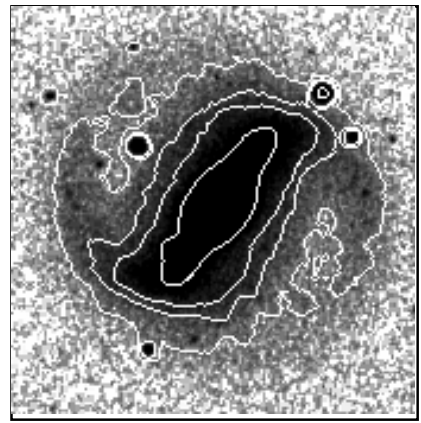}
\caption[]{DSS image of NGC~4314 in $B$.}
\label{dss4314}
\end{figure}

Boxy isophotes are a typical feature at the end of the bars of early type
(SB0, SBa) barred galaxies seen not far from face-on. Typical examples can be
found in Athanassoula, Morin, Wozniak et al. (1990) (NGC~936, NGC~4314,
NGC~4596), in Buta (1986) (NGC~1433), in Ohta, Hamabe and Wakamatsu (1990)
(NGC~2217, NGC~4440, NGC~4643, NGC~4665) and in many other papers. Loosely
speaking, the shape of these isophotes is rectangular-like. Their main
characteristic is that their small sides, at the largest distance of the
isophotes from the galactic center, are roughly parallel to the bar
minor axis (Athanassoula 1984, Athanassoula et al. 1990, Elmegreen
1996). In 
particular, Athanassoula et al. (1990) use generalized ellipses to fit
the isophotes of the bars in a sample of early-type strongly barred
galaxies, thus describing 
quantitatively the result that the shapes of the isophotes are
rectangular-like rather than elliptical-like.  Fig.~\ref{dss4314} is a
DSS image 
of NGC~4314 in $B$ and demonstrates a typical case of a galaxy with boxy
isophotes at the end of its bar.
We observe that the last isophotes of the bar are indeed rectangular-like.
Beyond the area of the boxy isophotes starts the spiral structure of
this galaxy.
 
There is a correspondence between the morphology of the boxy isophotes of the
early type barred galaxies and the isodensities encountered in snapshots of
several $N$-body models of bars. This was shown for the first time in
a simulation 
that was run specifically for this purpose (Athanassoula et al. 1990, Fig.~7).
In that paper this correspondence was underlined by measuring the
rectangularity in the same way as in the observations.  Since then there have
been several snapshots in $N$-body simulations reproducing this morphological
feature (see e.g. Shaw, Combes, Axon et al. 1993; Friedli \& Benz 1993;
Debattista \& Sellwood 2000). Recently Athanassoula \& Misiriotis (2002) in
their $N$-body models describe this feature also quantitatively. We
can thus conclude that both
observations and numerical models clearly show that boxiness close to the end of
the bars is a very frequently encountered phenomenon, and thus it should be
related to the standard dynamical behaviour in such systems.

Early calculations of orbits in 2D static potentials have underlined the
presence of rectangular-like periodic orbits at the 4:1 resonance region
(Athanassoula, Bienayme, Martinet et al. 1983; Contopoulos
1988; Athanassoula 1992a). They are either orbits of the x1 family on the
decreasing part of the characteristic -- towards lower $x$ values --
in type-2 4:1 resonance gaps\footnote{for the nomenclature of the gaps
see Contopoulos (1988), or Contopoulos \& Grosb{\o}l 
(1989).}, or orbits at the `4:1 branch' beyond the type-1 gap \cite{gco88}.
These orbits, whenever they exist, support outer boxiness on the face-on views
of the models. Their mere presence, however, is not sufficient to explain the
observed morphology. Explanations based on the presence of rectangular-like
planar 2D x1 orbits suffer from the following problems:
\begin{itemize}
\item The range of the Jacobi integral\footnote{We will hereafter
refer to the
    Jacobi integral as the `energy'.} over which one finds {\em stable}
  rectangular-like x1 orbits (in type-2 gaps) or 4:1 orbits (in type-1 gaps)
  is in general narrow (Contopoulos and Grosb{\o}l 1989). Furthermore,
  these orbits develop loops at 
  the four corners, whose size increases considerably with energy, while the
  near-horizontal sections of the orbit approach the minor axis (as in Fig.~3d
  in Athanassoula 1992a). This happens for energies only a little larger than
  the energy at which the orbits become rectangular-like.  Orbits with loops
  cannot easily reproduce the observed boxiness. Thus, we have only a very
  small energy interval with useful orbits (see Fig.~8
  in Athanassoula et al. 1990).
\item Poincar\'{e} sections for an energy value within the small energy
  interval where rectangular-like orbits are stable, show that
  the size of the stability area is very small (see e.g. Fig.~21 in Patsis et
  al. 1997a). This renders the trapping of many non-periodic
  orbits around stable boxy periodic orbits rather difficult. It is
  characteristic that Patsis, 
  Efthymiopoulos, Contopoulos et al. (1997b) used dynamical spectra in order
  to trace tiny stability islands of rectangular-like orbits in a 2D Ferrers
  bar potential.
  \item In many models the orbital rectangles are not sufficiently elongated.
  They frequently are rather square-like, even in cases of strong bars (as in
  Fig.~3c in Athanassoula 1992a).
\end{itemize}

Besides the papers of Contopoulos on non-linear phenomena at the 4:1
resonance region in 2D bars in the '80s, and the work of Athanassoula
and collaborators on the 
morphology of bar orbits in the early '90s, where the problem is explicitely 
stated, not much work has been done on this issue.
Nevertheless, from figures of orbits  in models of 3D bars (Pfenniger
1984; 1985) 
one can infer that problems like the squareness of the rectangular-like orbits
persist even if one considers also the third dimension of the bars. The
morphology of single orbits, however, does not determine the boxiness of the
isodensities of a model.

The goal of the present paper is to reconsider the problem of
rectangular-like isodensities in the framework of 3D orbital structure
models, and, more specifically, to examine the contribution of the
third dimension to the face-on orbital profiles. Based on the models presented
in Skokos, Patsis \&
Athanassoula (2002a,b; Papers I and II respectively), we investigate the
parameters which 
favour the presence of boxy outer isophotes in the face-on profiles of 
3D bars. As in most orbital structure studies, the models we present
are not self-consistent. As underlined in 
Papers I and II, our goal is to study the orbital behaviour, and
therefore the morphological changes, as a function of the model
parameters. For this we consider even extreme cases, in order to make
the effects clearest. We thus determine the main parameters that
influence the
boxiness of a model. We base our results not on the morphology of single
orbits, but on collective appearance when orbits of more than one
family are taken into account simultaneously. In parallel to this we study, as
counterexamples, cases where the 
appearance of rectangular-like orbits close to the end of the bar is excluded.
Finally, we discuss the correspondence between boxiness observed in edge-on
profiles and boxiness observed in the middle of the bars when viewed face-on.

\section{Face-on profiles}
In this series of papers (Papers I, II, and Patsis, Skokos \& Athanassoula
2002, hereafter Paper III) we study the basic families
in a general model composed of a Miyamoto disc of length scales A=3 and B=1, a
Plummer sphere bulge of scale length 0.4 and a Ferrers bar of index 2
and axial ratio $a:b:c = 6:1.5:0.6$. The masses of the three components
satisfy \( G(M_{D}+M_{S}+M_{B})=1 \) and are given in Table~\ref{tab:models}.
The length unit is 1~kpc, the time unit is 1~Myr and the mass unit is $ 2\times
10^{11} M_{\odot}$.  In the present paper we examine two additional models.
One of them, model B2, is characterized by a very strong bar, whose mass is
40\% of the total, and whose remaining parameters are as in model B (Paper
II). The other additional model has different axial ratios than the rest of
our model bars. It has $a:b:c=6:1:0.6$, instead of $a:b:c=6:1.5:0.6$, as all
others. It is used to study the contribution of families related to the
radial 3:1 resonance to the appearance of boxy isophotes at the end of bars.
The rest of its parameters are as in model A1, and we thus call it A1b. The
basic properties of the models we present in this paper are summarized in
Table~\ref{tab:models}.

A fundamental conclusion of Papers I and II is that essentially the backbone
for building 3D bars is the x1 tree of families of periodic orbits. The tree
consists of the x1 planar orbits and of its 3D bifurcations at the vertical
resonances. Only in one case (family z3.1s in model B) did we find a family
which supports the bar without being introduced in the system after a x1
bifurcation. In the present paper, we use these families in order to present
the face-on views of the skeletons of the models. We use for this purpose sets
of weighted orbits as in Paper III.  As we explained in Paper III, in order to
build a profile of weighted orbits for a model, we first calculate a set of
periodic orbits and pick points along each orbit at equal time steps. We keep
only stable representatives of a family. The `mean density' of each orbit (see
\S 2.2 of paper III) is considered as a first approximation of the orbit's
importance and is used to weight the orbit. We construct an image (normalized
by its total intensity) for each calculated and weighted orbit, and then, by
combining sets of such orbits, we construct a weighted profile. The selected
stable orbits are equally spaced in their mean radius. The step in mean radius
is the same for all families in a model.  We underline the fact that the
orbital profiles we present throughout the paper comprise only stable periodic
orbits.

As we have seen, the edge-on orbital profiles (paper III) are of stair-type,
which means that the families building the outer parts of the bars have the
lowest $\overline{|z|}$. We are thus here mainly interested in orbits
remaining close to the equatorial plane, since these orbits will contribute
more to the surface density at the end of the bars.
\begin{table*}
\caption[]{Parameters of our models. G is the gravitational constant,
M$_D$, M$_B$, M$_S$ are the masses of the disk, the bar and the bulge
respectively, $\epsilon_s$ is the scale length of the bulge, \O$_{b}$
is the pattern speed of the bar, $R_c$ is the corotation radius. The comment
in the last 
column characterizes the model  in order to facilitate
its identification.}
\label{tab:models}
\begin{flushleft}
\begin{tabular}{cccccccc}
model name& GM$_D$ & GM$_B$ & GM$_S$ & $\epsilon_s$ & \O$_{b}$ 
& $R_c$ & comments\\
\hline
A1 & 0.82 &  0.1  & 0.08 & 0.4 &  0.0540 &   6.13 &
fiducial \\                               
A2 & 0.82 &  0.1  & 0.08 & 0.4 &  0.0200 &  13.24 &
slow bar \\                               
A3 & 0.82 &  0.1  & 0.08 & 0.4 &  0.0837 &   4.19 &
fast bar \\                               
B  & 0.90 &  0.1  & 0.00 & --  &  0.0540 &   6.00 &
no bulge \\                               
C  & 0.82 &  0.1  & 0.08 & 1.0 &  0.0540 &   6.12 &
extended bulge \\                         
D  & 0.72 &  0.2  & 0.08 & 0.4 &  0.0540 &   6.31 &
strong bar \\                             
B2  & 0.60 &  0.4  & 0.00 & --  &  0.0540 &    6.51 &
no bulge/very strong bar \\                               
A1b & 0.82 &  0.1  & 0.08 & 0.4 &  0.0540 &    6.10 &
$a:b:c=6:1:0.6$ \\ 
\hline
\end{tabular}
\end{flushleft}
\end{table*}

\subsection{Model A1}
\begin{figure*}
\epsfxsize=17cm \epsfbox{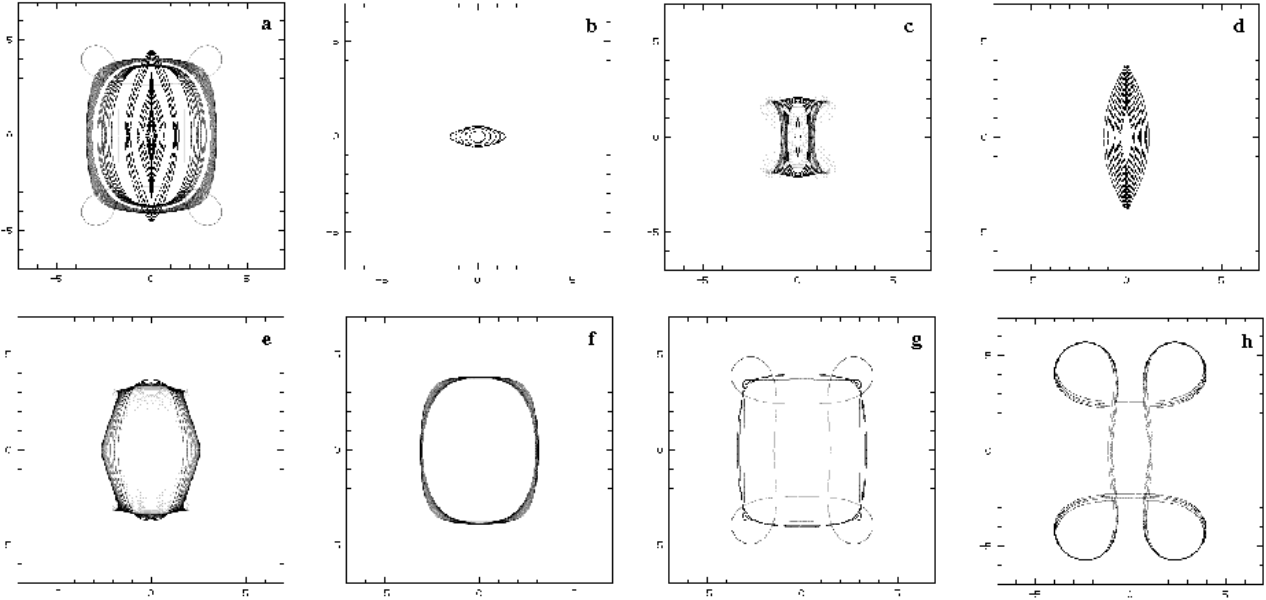}
\caption[]{The face-on, $(x,y)$, weighted profiles of the 3D families in model
A1. (a) x1, (b) x2, (c) x1v1, (d) x1v3, (e) x1v4, (f) x1v5,
(g) x1v7, and (h) x1v9.}
\label{A1xy}
\end{figure*}

The fiducial case model A1 offers, also for the face-on structure of the
models, a typical example of the contribution of the individual families to the
observed face-on orbital morphology. Fig.~\ref{A1xy} shows the weighted
profiles of all contributing families. It is evident by simple inspection 
that   
the orbits of the 2D family x1 (Fig.~\ref{A1xy}a) are the most important,
mainly because they have stable representatives over a large energy range.
Nevertheless, the projections of the 3D families depicted from
Fig.~\ref{A1xy}c to Fig.~\ref{A1xy}h play an important role. Contrarily, the x2
orbits (Fig.~\ref{A1xy}b) affect only the central parts of the system. As we
can see, boxy features are related to the families x1v1 (Fig.~\ref{A1xy}c),
x1v4 (Fig.~\ref{A1xy}e), x1v5 (Fig.~\ref{A1xy}f), x1v7 (Fig.~\ref{A1xy}g) and
x1v9 (Fig.~\ref{A1xy}h). The family x1v3 (Fig.~\ref{A1xy}d), on the other hand,
has always orbits with elliptical-like projections, thus in a way plays a
complementary role, together with x1, providing building blocks for
the elliptical-like part of the bar.

The orbits with the rectangular-like projections in model A1 face one of the
main problems for explaining the boxy isophotes, i.e. they are less elongated
than necessary. In general the boxy isophotes of the early type bars are not
squares. 
The rectangular-like orbits in model A1 have also
shorter projections than the x1 ellipses on the semi-major axis. An exception
is x1v9 (Fig.~\ref{A1xy}h). However, the four loops of these orbits are larger
than can be admitted by the corresponding shapes of the isophotes of real
galaxies
Furthermore, the x1v9
family contributes, because of its stability, only over a narrow energy
interval ($-0.185 < E_j < -0.182$).

It is interesting to note that the boxy x1v1 orbits, responsible in many
models for peanut-shaped edge-on profiles, provide to the face-on view of the
system a `bow tie' structure. However, in this particular case at least, these
orbits remain confined well inside corotation and can be responsible only for
an inner boxiness in a galaxy and not for boxy isophotes at the end of the
bars.

\begin{figure}
\hspace{1cm}\epsfxsize=7.0cm \epsfbox{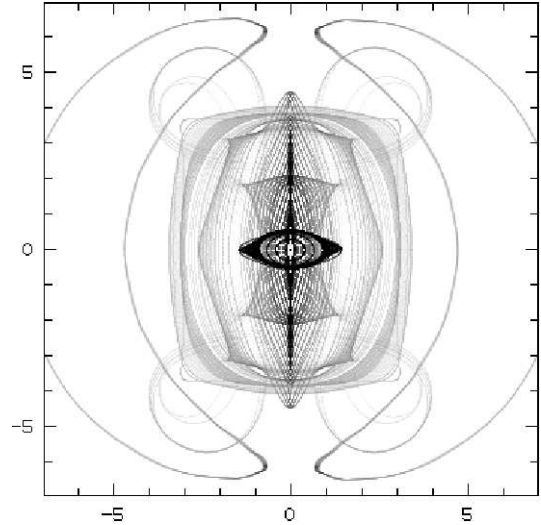}
\caption[]{Face-on orbital profile for model A1. All  x1-related
  orbits and the banana-like orbits are included, and they are  weighted
  as described in paper III.}
\label{A1all}
\end{figure}

Fig.~\ref{A1all} combines all x1 related families, i.e. the 2D x1 family and
its 3D bifurcations, as well as the x2 and the long-period banana-like orbits.
We can see that the orbits support a bar with a semi-major axis of length
about 0.75 of the corotation radius, so that the ratio of the corotation
radius $R_c$ to the orbital length of the bar $a_o$ will be $R_c/a_o = 1.33$.
The longest orbits along the bar are elliptical-like. This however, is
not an obstacle to forming more rectangular-like bars, since the
elliptical-like orbits extend only little beyond the rectangular-like ones.
This extra extent could be suppressed if the outermost periodic elliptical-like
orbits were not populated, or it could be masked by orbits trapped around the
periodic rectangular-like ones. In the latter case, the isophotes (or rather
isodensities) would be more elongated than the corresponding rectangular-like
orbits.

We note that in a galaxy or in an $N$-body simulation, not all families
included in the figure should be necessarily populated. In our
non-self-consistent models we try to identify structures in order to seek
their corresponding features in galactic images and snapshots of
self-gravitating models.

\subsection{Model A2}
\begin{figure*}
\epsfxsize=18.0cm \epsfbox{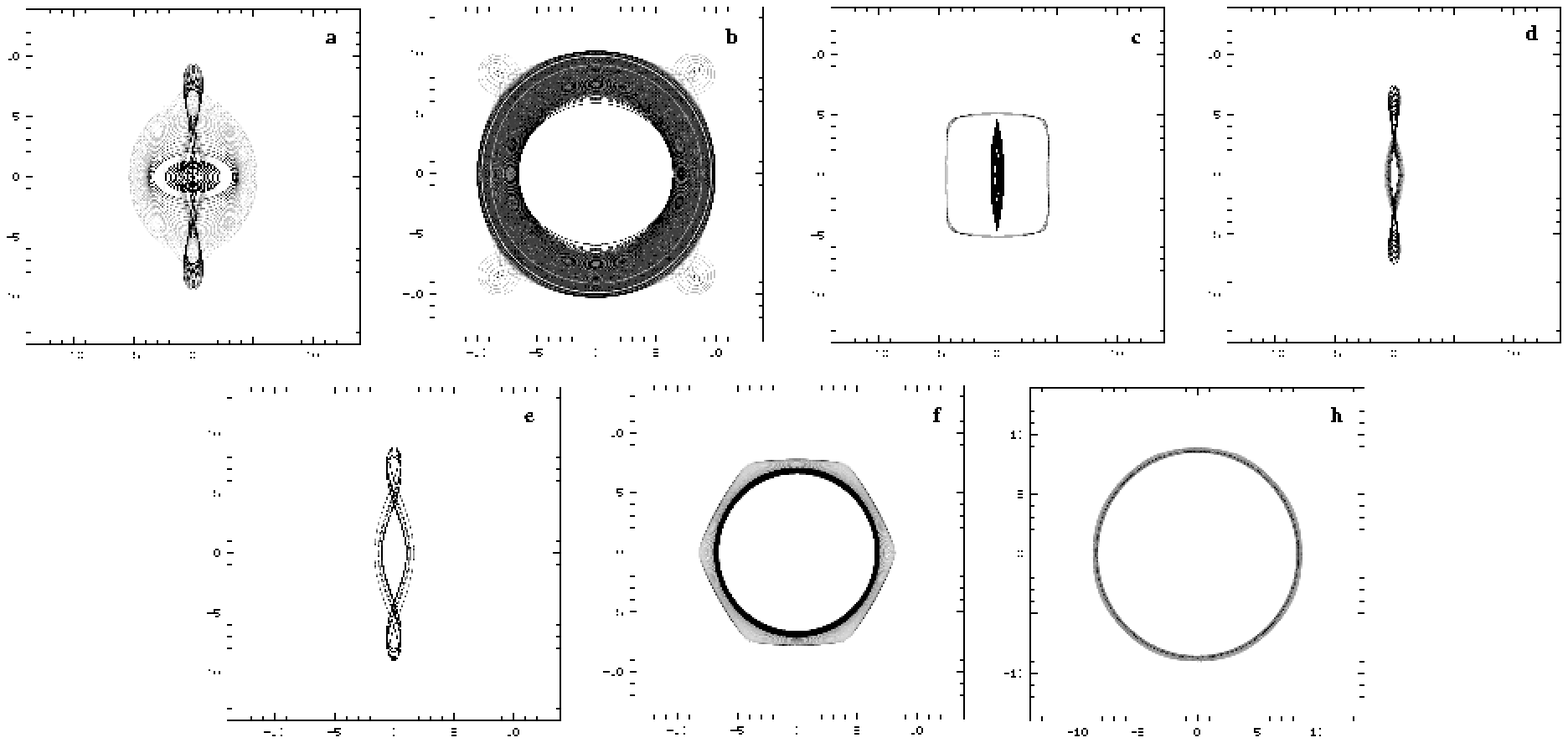}
\caption[]{The face-on, $(x,y)$, profiles of the 3D families in model
A2. (a) x1 and x2, (b) x1$^{\prime}$, (c) x1v1, (d) x1v3, (e) x1v4, (f)
x1$^{\prime}$v4, and (h) x1$^{\prime}$v5. }
\label{A2xy}
\end{figure*}

The slow rotating bar in model A2 brings new morphological features. In this
case square- or rectangular-like orbits play a minor role, as we realize from
Fig.~\ref{A2xy}. Only x1$^{\prime}$ orbits (Fig.~\ref{A2xy}b) at the
decreasing branch of the characteristic (paper II) and part of the x1v1
family, after the
S\ar \D transition (Fig.~\ref{A2xy}c), provide such stable orbits to the
system. For both these families, however, the size of the orbits and the
energy range over which they exist do not allow them to play a major role in
the orbital structure of the model. In Fig.~\ref{A2xy}b we choose the contrast
of the image such as to allow us to see the loops of the x1$^{\prime}$
square-like orbits.  For even larger energies the loops of the x1$^{\prime}$
square orbits become huge, and finally the orbits become retrograde. Such
orbits are not depicted in Fig.~\ref{A2xy}b.

\begin{figure}
\epsfxsize=8.5cm \epsfbox{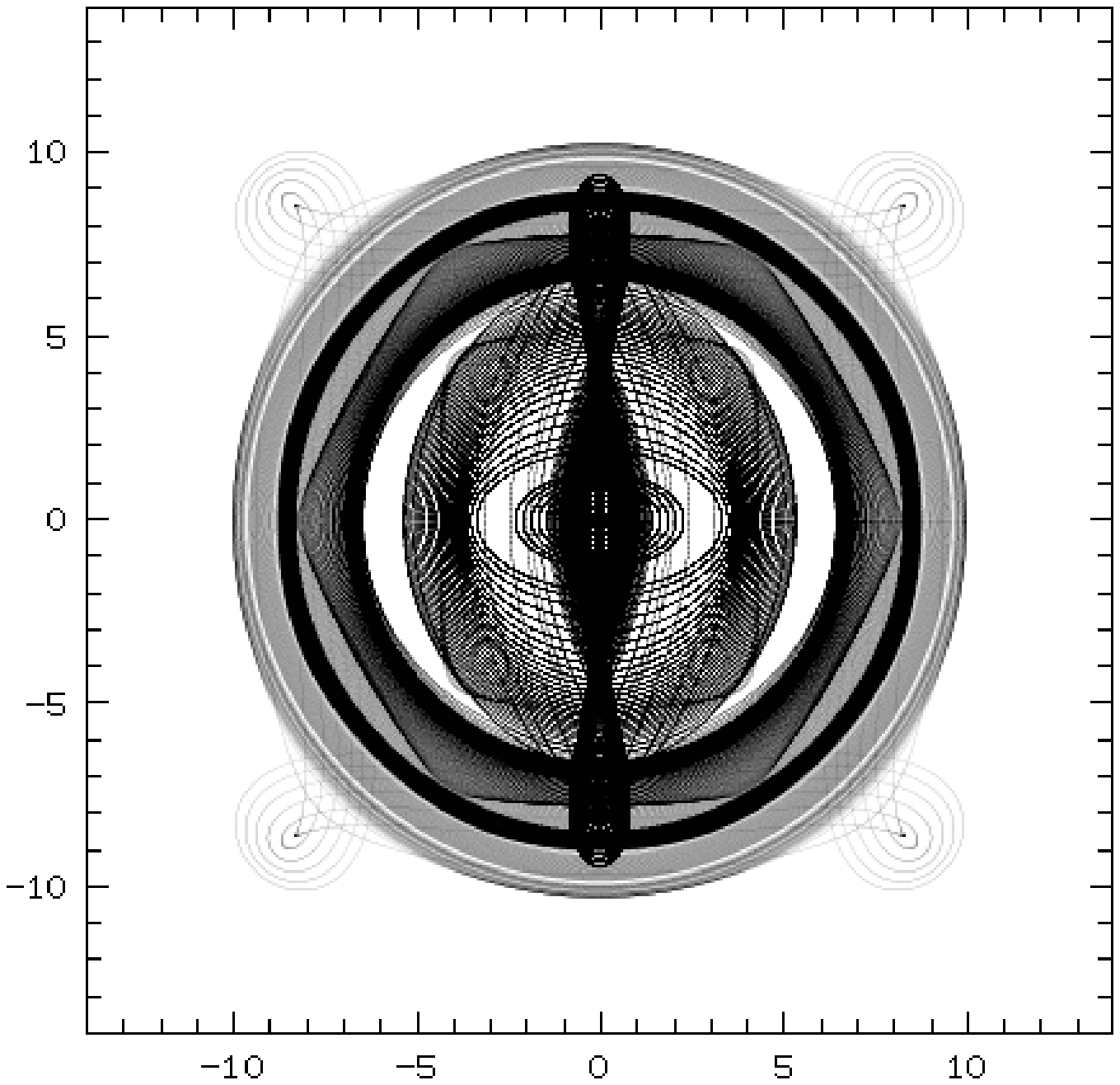}
\caption[]{Face-on orbital profile for model A2. All  
  orbits are weighted  as described in paper III.}
\label{A2all}
\end{figure}

Model A2 has two main features: First, the loops along the bar major axis,
which are brought in the system by the x1 family and its x1v3 and x1v4
bifurcations. Second, the almost circular (and/or square-like) projections on
the equatorial plane of the families x1$^{\prime}$, x1$^{\prime}$v4, and
x1$^{\prime}$v5 (Fig.~\ref{A2xy}b, f, h respectively). The face-on morphology
supported by this slow rotating bar is depicted in Fig.~\ref{A2all}.
In this figure we combine all orbits of the families presented in
Fig.~\ref{A2xy}. The result is a bar with loops along the major axis
surrounded by almost circular orbits. We note that the bar-supporting orbits
extend to a distance about 9 from
the center (corotation in this case is at 13.24), i.e. in this case $R_c/a_o =
1.5$. 

\subsection{Model A3}

\begin{figure*}
\epsfxsize=14.0cm \epsfbox{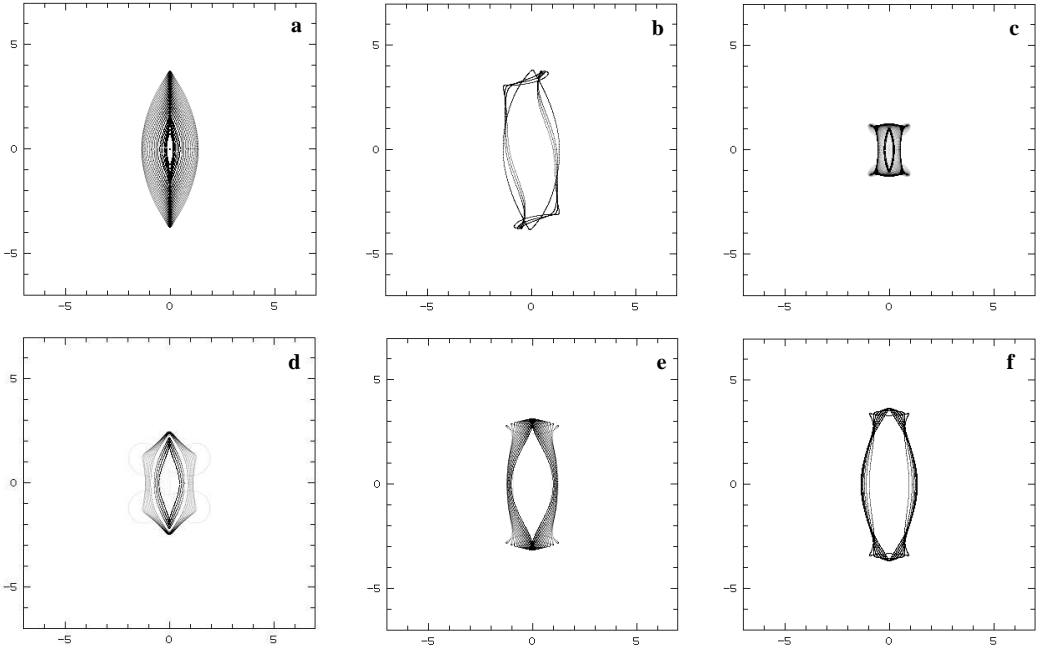}
\caption[]{The face-on, $(x,y)$, profiles of the 3D families in model
A3. (a) x1, (b) q0, (c) x1v1, (d) x1v3, (e) x1v5, (f) x1v8.
}
\label{A3xy}
\end{figure*}

If the bar rotates fast (model A3), the face-on orbital structure changes
significantly. The x1 family dominates once again, as we can see in
Fig.~\ref{A3xy}a. Its orbits remain always elliptical-like, but now they
support a bar with length 0.95 of the corotation radius, i.e. $R_c/a_o =
1.05$.  The dynamics at the radial 4:1 resonance region are crucial for the
morphology of the model (paper II). In this case the families q0 and x1v8
provide the system with rectangular-like orbits quite elongated along the
major axis of the bar. Family q0 can be found in two branches,
symmetric with respect to the bar's major axis. We can, however,
consider orbits of only one of its two branches if we want in our 
weighted profiles a non-rectangular parallelogram-like morphology. In order to
obtain a desired 
morphology in our response models, we chose any combination of
stable orbits from all available families.
In the present case both families q0 and x1v8 have orbits at least as long as
the x1 orbits. In addition, families x1v1 and x1v5 also have orbits with boxy
projections on the equatorial plane, but reaching distances from the center
1.2 and 3.2 respectively (Fig.~\ref{A3xy}c,e). That means that, if all boxy
families are populated, we can have boxy isophotes in several scales in this
model. Even the hexagonal-like projections of the orbits of family x1v3
(Fig.~\ref{A3xy}d) contribute to the boxiness of the model with their sides
which are parallel to the bar major axis.
The face-on view of this family has also the `bow tie' appearance.

\begin{figure}
\epsfxsize=6.5cm \epsfbox{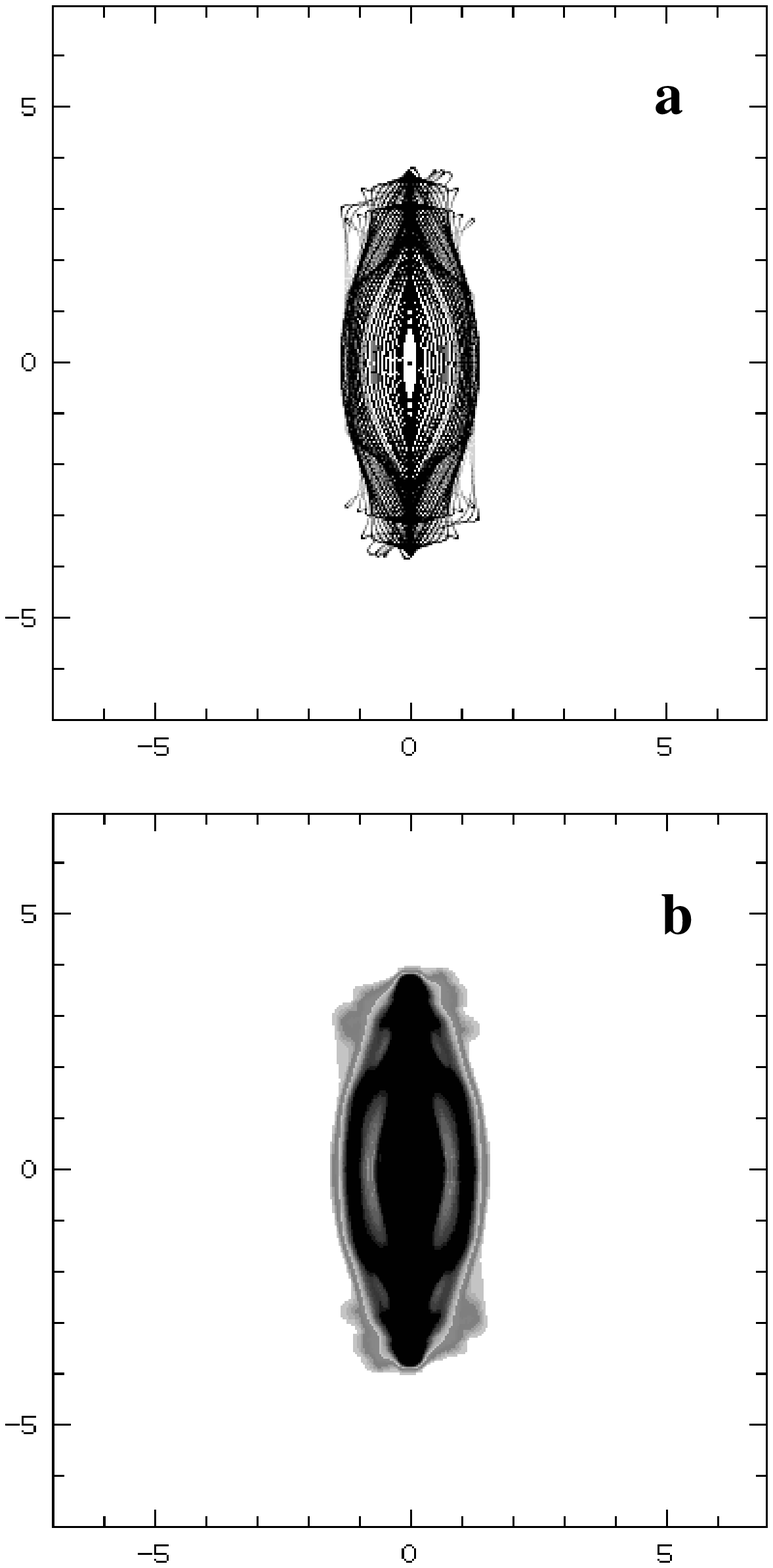}
\caption[]{(a) Composite profiles combining all orbits in model A3. (b) The
  profile after applying a gaussian filter in order to get an impression of
  the morphological features supported by the orbits.}
\label{A3all}
\end{figure}

The total effect when considering all orbits is given in Fig.~\ref{A3all},
where we take into account all families of Fig.~\ref{A3xy}. In
Fig.~\ref{A3all}a we give all weighted orbits together, while in
Fig.~\ref{A3all}b we apply a gaussian filter in order to show clearly, in a
first approximation, the shapes of features, which could be supported by the
orbits in the density maps of the models. Fig.~\ref{A3all} tells us that the
fast rotating bar model, has a boxy bar with $R_c/a_o \approx 1.05$.

The fast rotating bar case offers the opportunity to study models with a
substantial non-axisymmetric force near corotation. In this way we compensate
for the standard shortcoming of Ferrer's \footnote{ We use Ferrers bars
  because we are not aware of any other models which are {\em realistic} and
  {\em analytic}, and do not have this shortcoming.}  ellipsoids, namely that
their force drops too fast as the radius increases and approaches corotation.
Thus A3 allows us to cover also cases with substantial non-axisymmetric
forcing near corotation.

\subsection{Model B}
\begin{figure*}
\rotate[r]{
\epsfxsize=4.20cm \epsfbox{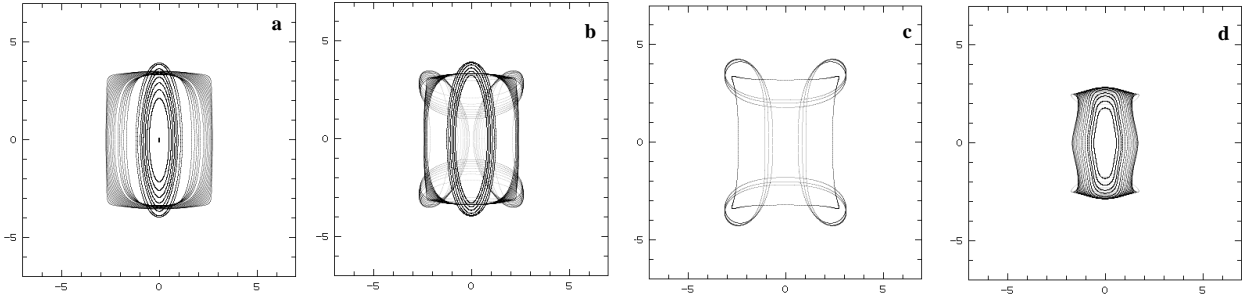}}
\caption[]{The face-on, $(x,y)$, profiles of the 3D families in model
B. (a) x1, (b) x1v5 and x1v5$^{\prime}$, (c) x1v7, (d) z3.1s }
\label{Mxy}
\end{figure*}

Model B is a model without radial or vertical 2:1 resonances. As we have seen
in paper II, the first vertical bifurcation of x1 is x1v5. Thus, the families
that build the bar are x1, x1v5/x1v5$^{\prime}$, x1v7 and the family z3.1s.
We remind that this latter family plays a significant role in the orbital
behaviour of this particular model (see Paper II) and it was found as a
bifurcation of the z-axis orbits when the latter are considered as being of
multiplicity 3, i.e. the orbits are repeated three times. The z3.1s family is
not related to the x1-tree. The weighted face-on profiles of the above
mentioned families are given in Fig.~\ref{Mxy}. We see that all of them have
boxy representatives in their projections on the equatorial plane.

\begin{figure}
\epsfxsize=7cm \epsfbox{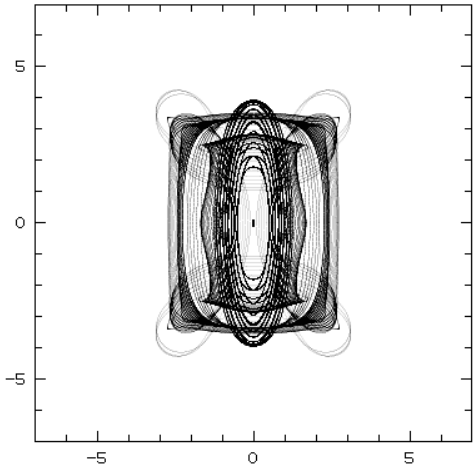}
\caption[]{ Composite profile combining the orbits of families x1,
  x1v5/x1v5$^{\prime}$, x1v7 and z3.1s  in model B.
  } 
\label{Mall}
\end{figure}

In Fig.~\ref{Mall} we have the orbital face-on view for model B obtained by
overplotting the weighted orbits of all families together. In this model 
$R_c/a_o \approx 1.4$. 

\subsection{Model D}
\begin{figure*}
\epsfxsize=14.0cm \epsfbox{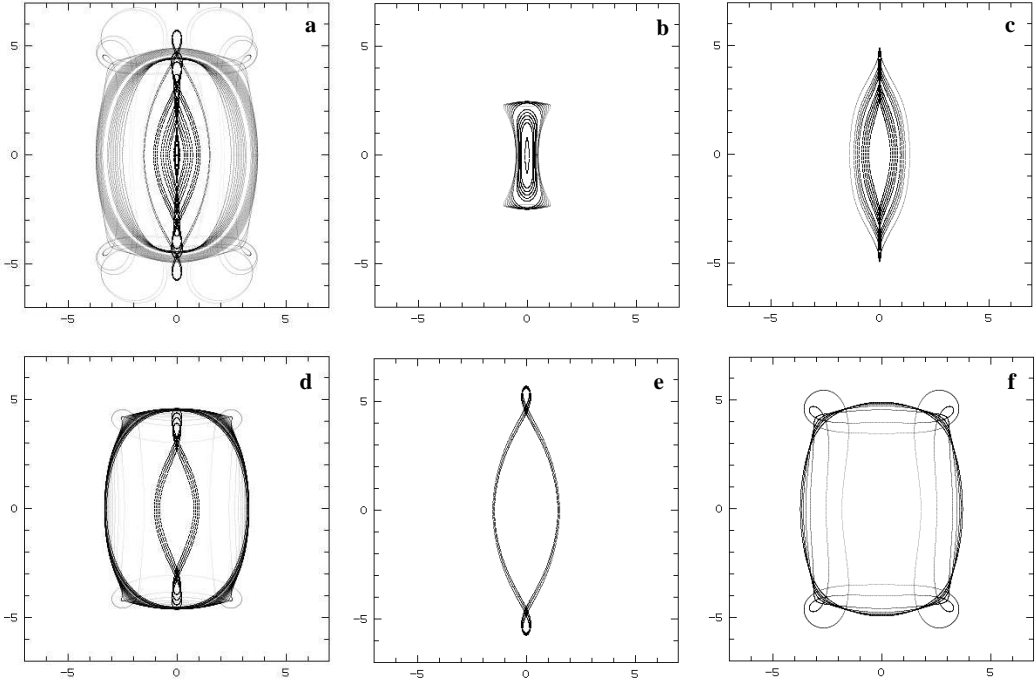}
\caption[]{The face-on, $(x,y)$, profiles of the 3D families in model
D. (a) x1, (b) x1v1, (c) x1v3, (d) x1v5/x1v5$^{\prime}$, (e) x1v6, (f) x1v7
 }
\label{Hxy}
\end{figure*}

The strong bar model D is, after the fast rotating bar model A3, the second
case where we have periodic orbits that can contribute to sufficiently
elongated rectangular-like boxy isophotes. They are, however, considerably
less elongated than the rectangular orbits in A3. As we see in Fig.~\ref{Hxy},
the main contributors are now families x1, x1v5/x1v5$^{\prime}$ and x1v7.
Again x1v1 orbits have `bow tie' face-on projections, and support a feature of
corresponding morphology (Fig.~\ref{Hxy}b). This feature occupies the main
part of the bar, but it does not reach its end, which is at about a distance
0.9 of the corotation radius.

\begin{figure}
\epsfxsize=7cm \epsfbox{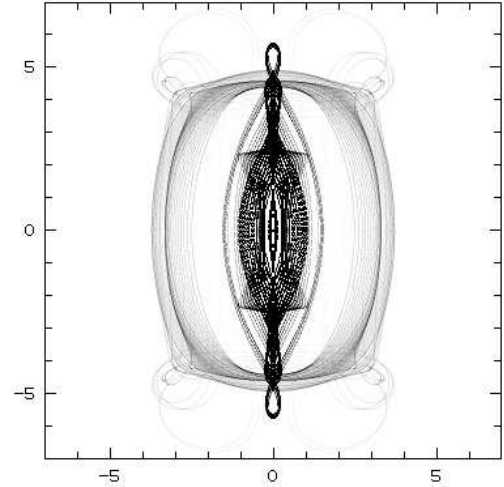}
\caption[]{Composite face-on profile of model D. All orbits of Fig.~\ref{Hxy}
  are considered.}
\label{Hall}
\end{figure}

The composite face-on orbital profile of model D is given in Fig.~\ref{Hall}.
We see that --although the boxy orbits are present and, if populated, could
characterize the morphology of the model-- the dominant feature is the loops
of the x1 orbits along the major axis. This morphology is enhanced by the
loops of the projections of the 3D bifurcations of x1, x1v5/x1v5$^{\prime}$
and x1v6 (Fig.~\ref{Hxy}). It should be noted, however, that only the loops of
x1v6 and some of the x1 loops extend beyond the rectangular outline. If we
omit orbits with such loops we get, for this model, $R_c/a_o \approx 1.23$, 
while if we include them we get $R_c/a_o \approx 1.08$.

\subsection{Very massive bars}
\begin{figure*}
\rotate[r]{ 
\epsfxsize=7.5cm \epsfbox{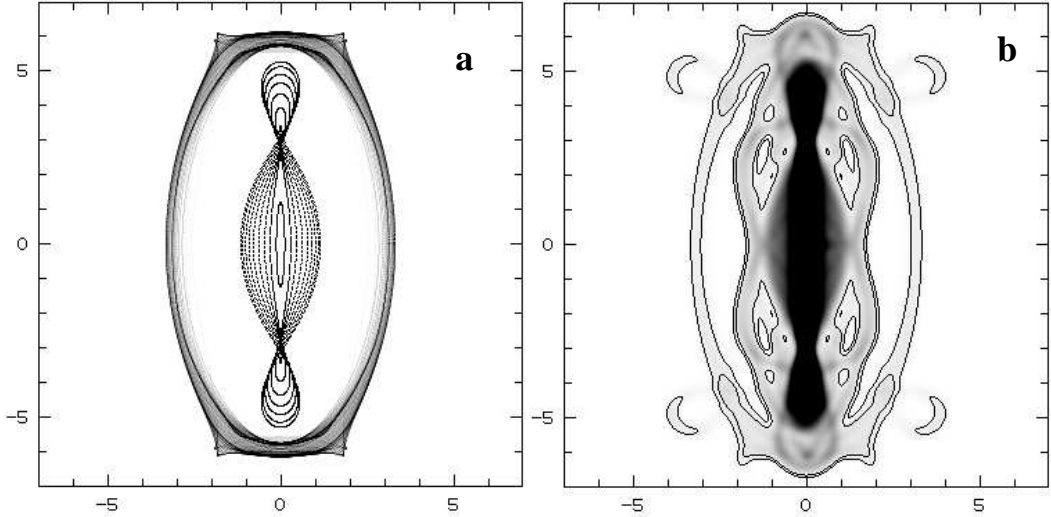}
}
\caption[]{(a) A composite face-on profile of model B2. We consider orbits
  from the
  families x1, x1v5/x1v5$^{\prime}$and x1v7. (b) The blurred profile of model
  B2. Besides the x1, x1v5/x1v5$^{\prime}$and x1v7 we consider stable orbits of
  x1v6 and 2D orbits of family t1. In model B2 we can have rectangular-like
  face-on profiles with $p_{max}/p_{min}$ ratio close to 2.
  }
\label{modelB2}
\end{figure*}

In all the models we studied so far the rectangular-like orbits, if they
exist, are rather square-like, except for the fast rotating bar case (model
A3), where they are more elongated. The second parameter, after the pattern
speed, which made the rectangular-like orbits more elongated is the increase
of the relative mass of the bar. For this reason, we studied several more
models with increased bar mass fraction and calculated the orbital stability
of each individual case. We stopped when we reached models with very large
intervals of instability of the x1 family and its 3D bifurcations.  From a
sequence of models starting with the values of the parameters of model B and
with an increasingly large fraction of mass in the bar, we found that this
fraction is about 50\% of the total mass. It was evident that models with bar
masses about 40\% of the total mass offered families of periodic orbits with
enough stable representatives to support sufficiently elongated
rectangular-like isophotes at the end of the bars. E.g. model B2, which
differs from model B only in that the mass of the bar GM$_B$=0.4 instead of
0.1 as in model B, has rectangular-like orbits with the ratio of their
projections on the major axis of the bar to their projections on the minor
axis ($p_{max}/p_{min}$) close to 2. This can be seen in Fig.~\ref{modelB2}.
In Fig.~\ref{modelB2}a we have orbits from the families x1,
x1v5/x1v5$^{\prime}$ and x1v7, omitting rectangular-like orbits with loops.
The blurred image in Fig.~\ref{modelB2}b includes also orbits of the t1 family
(see Paper I) and stable orbits of family x1v6. The latter is initially
bifurcated as unstable. For larger energies, however, it has stable orbits
with face-on projections with loops along the major axis of the bar.  The t1
orbits, bifurcated at the radial 3:1 resonance, introduce in the system some
interesting features if one considers both their branches which are symmetric
with respect to the minor axis of the bar at a given energy. The t1 orbits are
responsible for the guitar-like feature we observe in Fig.~\ref{modelB2}b
closer to the major axis than the outer rectangular-like orbits. 
For model B2 we have $R_c/a_o \approx 1.04$.

\subsection{Support by 3:1 families}
\begin{figure}

\epsfxsize=7.5cm \epsfbox{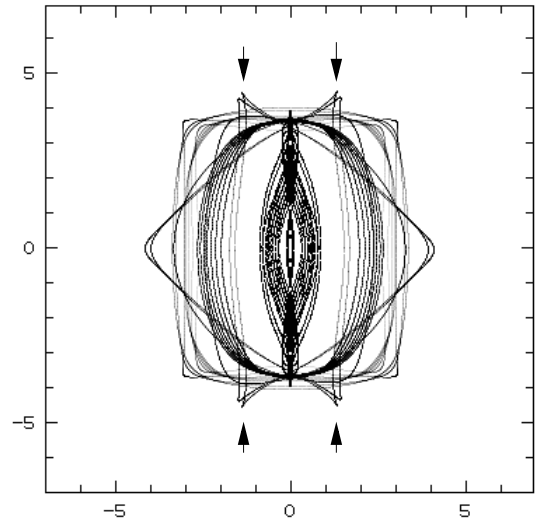}

\caption[]{x1 and t2 orbits in model A1b. The elliptical-like orbits with
  loops along the major axis of the bar are clearly separated from the rest of
  the orbits. Due to orbital instabilities the boundary between the two regions
  has a rectangular-like shape and is enhanced by the t2 orbits. Its four
  corners are indicated by arrows. 
  }
\label{t2orb}
\end{figure}

The contribution of t1 to the features we found in model B2 gave us the
incentive to investigate the contribution of families bifurcated at the radial
3:1 resonance in all our models. We found that in general orbits from the t1
family may support motion roughly parallel to the minor axis of the bar, but
not at the end of the bar. Fig.~\ref{modelB2}b describes the kind of
contribution t1 orbits can offer to the overall boxiness of face-on profiles
of barred galaxies. We give also an example of the combination of t2 orbits 
(see Paper I) in
a model in which this contribution is pronounced. It is model A1b, a model
that differs from the fiducial case (model A1, Paper I) only in the axial
ratios, which in model A1b are $a:b:c=6:1:0.6$ instead of $a:b:c=6:1.5:0.6$.
In Fig.~\ref{t2orb} we observe
stable x1 orbits with loops along the major axis of the bar, and x1 orbits
with a more or less rectangular like shape. The two groups are separated by an
empty region caused by an instability region of the x1 family. On top of the
x1 orbits we overplot stable orbits of both branches, symmetric with respect
to the major axis of the bar, of the t2 family. The t2 orbits contribute to
the boxiness of the model by enhancing the borders of the empty region and by
forming a rectangular-like feature indicated in Fig.~\ref{t2orb} by four black
arrows. Model A1b has $R_c/a_o \approx 1.35$.

\section{Discussion}
 We examined the face-on views of all models of Papers I
and II and we found families that could support outer, as well as inner, boxy
isophotes. In the present paper we do not refer explicitly to model C, since
its orbital behaviour does not differ essentially from that of model A1.
Furthermore, we use composite profiles of two additional models to demonstrate
the effect of very massive barred components and the possible role of families
related to the radial 3:1 resonance.

The first quantity that can be used to compare the bars of our models with the
bars of real galaxies concerns the length of the bar built by the orbits with
respect to corotation radius. Indeed observations (for a compilation
see e.g. Athanassoula 1992b,
Elmegreen 1996 and Gerssen 2002) as well as hydrodynamical
simulations (Athanassoula 1992b), have shown that the ratio $R_c/a_o$ takes
values only in a restricted range, namely $R_c/a_o = 1.2 \pm 0.2$. This is in
agreement with 2D orbital models by Contopoulos (Contopoulos 1980). All our
response models have $R_c/a_o$ ratios within this range, except for model A2
which has 
a slow rotating bar and $R_c/a_o = 1.5$. Table~2 summarizes the $R_c/a_o$
\begin{table}
\caption[]{The ratio of the corotation
radius $R_c$ to the orbital length of the bar $a_o$ ($R_c/a_o$) for all models
studied in the present paper. We find $1.04 \lid R_c/a_o \lid 1.5$. The third
column gives a short comment about what characterizes the model  in order to
facilitate its identification.}
\label{2}
\begin{tabular}{ccc}
model name& $R_c/a_o$ & comments\\
\hline
A1  & 1.33 &  fiducial \\                               
A2  & 1.50 &  slow bar \\                               
A3  & 1.05 &  fast bar \\                               
B   & 1.40 &  no bulge \\                               
D   & 1.08 &  strong bar \\                             
B2  & 1.04 &  no bulge/very strong bar \\                               
A1b & 1.35 &  $a:b:c=6:1:0.6$ \\ 
\hline
\end{tabular}
\end{table}
ratios for the models we studied in the present paper.
 
A second quantity we can compare with galaxies and snapshots of $N$-body
models is 
the ratio $p_{max}/p_{min}$ for the rectangular-like orbits. Eye estimates
show that the boxy outer isophotes at the end of the bar of strongly barred
galaxies have a ratio $p_{max}/p_{min}$ typically larger than 2 (e.g. NGC~936,
NGC~4314, NGC~4596). Exceptional cases can be found in the literature but are
not many (see e.g. the not-rectified images of NGC~1415 in Garcia-Barreto \&
Moreno (2000)). 
A more quantitative study has been made by Athanassoula et al (1990). They find
that the axial ratios of the isophotes near the end of the bar are, for all
the galaxies in their early-type strongly-barred galaxy sample, considerably
larger than 2. On the other hand, in most of our models the boxy
orbits are quite square-like. 
Ratios of boxy isophotes larger than 2 have been mainly found in the fast
rotating bar case, where $p_{max}/p_{min} \approx 2.5$. The strong bar model D
has $p_{max}/p_{min} \approx 1.4$, while in model B2, where the mass of the
bar is 40\% of the total mass, we attained a $p_{max}/p_{min}$ ratio close to
2. In the rest of the models the ratio was less than 1.4.

\begin{figure}
\epsfxsize=7cm \epsfbox{ 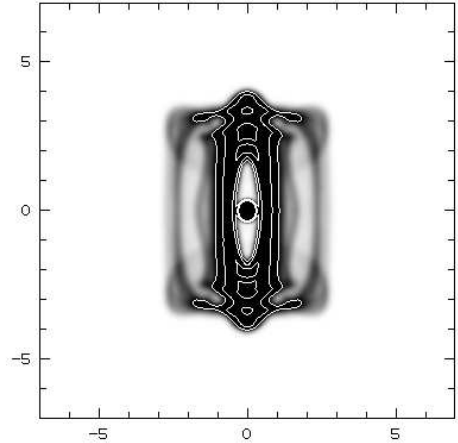}
\caption[]{Isodensities on a blurred image of model B. The narrow
  rectangular-like structure is supported by orbits of more than one family.}
\label{isobl}
\end{figure}

Let us underline that, for studying the elongation of the rectangular-like
isophotes observed in barred galaxies, one needs to combine orbits of several
families instead of evaluating the properties of a single orbit or a single
family. This is a result of the three dimensional character of our models and
of the fact that we have bars built by orbits belonging mainly to the
x1-tree. This 
effect is indicated by isodensities we plotted on some blurred images of our
models. Such a typical case is given in Fig.~\ref{isobl}, for model B.
The small sides of the rectangular-like structure outlined by the isophotes
reflect mainly the contribution of the rectangular-like orbits. However, the
two large sides, parallel to the major axis of the bar, are due to the
overlapping of orbits of the x1, and z3.1s families.

Our models show that it is the pattern speed that mainly
determines the elongation of the outer boxy orbits,. The
orbits building the bar of model A3 are rectangular-like and could make a bar
with the geometry encountered in the early-type bars with outer boxy
isophotes. It is just the increase of $\Omega_b$ that brought in the system
the 2D family q0 and the 3D, stable x1v8 orbits. We note the morphological
similarity of the q0 orbits with the outer boxy isophotes of NGC~4314
(Quillen, Frogel \& Gonzalez 1994), which have the form of a non-rectangular
parallelogram (cf.  Fig.~\ref{A3all}b with Fig.~\ref{dss4314}).
We note that in model A3 the rectangular-like bar ends closer to corotation
than the bars of the models for which the longest orbits are the x1 with loops
along the major axis. This is consistent with the result found in Patsis et
al. (1997a) for the pattern speed of NGC~4314, in which the boxy isophotes are
very close to the corotation radius indeed. It is  also obvious that, apart
from the increase of the pattern speed, the increase of the strength of the
bar favoured the elongation of the rectangular-like orbits.

Model A2, which is the slowest rotating case, is the other extreme of the
face-on morphologies we encounter in our models. In this model, planar boxy
orbits and projections of orbits on the equatorial plane are square-like, and
very little elongated along the major axis of the bar (Fig.~\ref{A2xy}c of the
present paper, and Fig.5 in paper II). The development of loops at the
apocentra of the x1 elliptical-like orbits, as well as in the projections on
the equatorial plane of the x1v3 and x1v4 families, together with the almost
circular x1$^{\prime}$ orbits and the projections of the x1$^{\prime}$v4 and
x1$^{\prime}$v5 families, give to model A2 a characteristic face-on morphology
with the circular-like orbits surrounding all other orbits in the face-on
projection. Such a morphology could be linked to the existence of inner rings
in barred galaxies.

Inner boxiness of the face-on profiles, much closer to the center than the
corotation region, is associated mainly with the x1v1 orbits, i.e. with the 3D
family born at the vertical 2:1 resonance. As we have seen in paper III, in
several models this family is responsible for peanut-shaped orbital structures
in the edge-on views of the models. The face-on orbital skeletons of the
models we present here show in another clear way what we noticed in paper III,
namely that the x1v1 family builds boxy- or peanut-shaped features which do
not approach corotation. Inner boxiness is not rare in galactic bars. Typical
examples of boxy isophotes in the middle of the stellar bars are NGC~3992 and
NGC~7479 (Wilke, M\"{o}llenhoff \& Matthias 2000). The x1v1 orbits,
or at least their representatives with the largest energy values (see
Fig.~\ref{A1xy}c, Fig.~\ref{A3xy}c and Fig.~\ref{Hxy}b), are boxy in
their face-on views, but they also have a characteristic `bow-tie'
morphology. Inner boxiness is 
also supported by the z3.1s orbits, at large values of 
the energy, in model B.

\begin{figure}
\epsfxsize=7cm \epsfbox{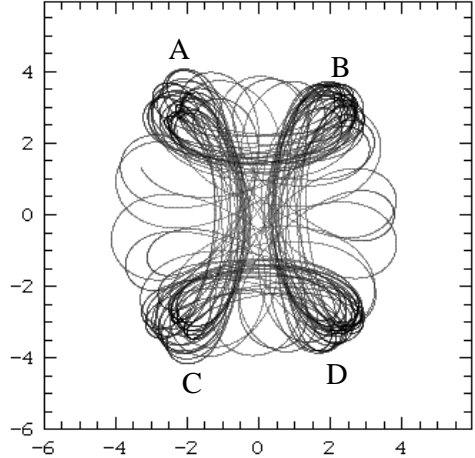}
\caption[]{An example of a chaotic orbit in Model B, that, for a small number
  of dynamical times, supports a `bow-tie'
  morphology. A particle following this trajectory spends most of the time in
  the regions of the loops indicated with A, B, C and D, while the `arcs'
  $\stackrel{\frown}{AC}$ and $\stackrel{\frown}{BD}$ approach the center of
  the galaxy. Darker areas indicate regions where the orbit spends more time.
  }  
\label{bowtie}
\end{figure}

A `bow-tie' morphology in  models of the kind we study here can also be
introduced by rectangular-like orbits at high energies. These orbits (or more
precisely their face-on projections) develop loops, which, in some cases, stay
close to the orbits that build the bulk of the bar. This is e.g. the case of
the x1v5$^{\prime}$ family in model B (Fig.~\ref{Mxy}b).  When the
rectangular-like orbits develop loops they also become of `bow-tie' shape.
Since their deviations from the equatorial plane increase with energy, their
projections occupy almost the same area as the rectangular-like orbits of the
same family, for lower energies, do. In model B, for $E_J \approx -0.185$,
integrating even chaotic orbits for a {\it small} number of dynamical times
(of the 
order of 30) we get a bow tie morphology. This is shown in
Fig.~\ref{bowtie}.
Non-periodic elliptical-like orbits trapped around the x1 family 
not far from the center of the bar, as well as the projection of
orbits trapped around the three-dimensional x1 bifurcations would fill
more densely the area between the regions A and B (and symmetrically
between C and D), than between the regions A 
and C (and B and D). This results from the elliptical-like shape of the
projections of these orbits on the equatorial plane and the fact that they
have their apocentra between the regions A and B (and C and D). We
note also the 
contribution of the x1v1 orbits to a `bow-tie' morphology of face-on profiles,
as mentioned previously.

Due to the presence of these families a bow-tie morphology, like the one found
by Barnes \& Tohline (2001), is not excluded from the usual barred morphology.
Nevertheless, we did not find orbits of a single family that can both enhance a
rectangular-like structure at the end of the bar without loops at its four
corners, and simultaneously enhance a bow-tie morphology. Single families with
bow-tie profiles in their face-on projections do not extend to the end of the
bar (x1v1 in Figs. 2c, 6c, 10b and x1v3 in Fig. 6d).

\section{Conclusions}
In this paper we discussed boxiness in the face-on views of 3D models, and, in
general, their face-on orbital structure. We examined a large variety
of possible 3D orbital behaviour, that could contribute to the
boxiness in face-on views of barred galaxies. Our main conclusions are:

\begin{enumerate}
\item Boxiness in the face-on views of 3D barred models is an effect
  caused by the co-existence of several families, each contributing
  appropriate  stable orbits. The
  morphology of boxy isodensities/isophotes is not necessarily similar to the
  morphology of individual stable, rectangular-like orbits. In some cases
  (Fig.~\ref{isobl}) the iso-contours can be narrower than the orbits.
\item In models with the right morphological
  parameters, we find appropriate building blocks to account for the
  rectangular-like isophotes or 
  isodensities seen in early type barred galaxies and in some $N$-body
  simulations. 
\item In 3D models the family of the planar x1 orbits is subject to vertical
  instabilities, and thus in several cases it has considerable instability
  strips at the 4:1 resonance region.  This should be an
  obstacle for x1 planar orbits to account for the rectangular-like shape of
  bars, since unstable periodic orbits can not trap regular orbits around
  them. However, at the instability regions of the x1 we find other stable
  families, whose ($x$,$y$)-projected orbital shapes are, at least near their
  bifurcations, very similar to those of the x1. They have orbits which are
  stable over large energy intervals and also have ($x$,$y$)-projected
  shapes that 
  can enhance a rectangular-like bar outline.  Thus the inclusion of the third
  dimension in the models enhances the possibility of rectangular-like
  isodensities.
\item There are families of periodic orbits that support boxiness in
  the outer bar regions, as well as families that support boxiness in
  somewhat more inner parts.
  The standard families belonging to the former category are the stable
  representatives of the x1 orbits (close to the radial 4:1 resonance), and of
  the families x1v5, 
  x1v7 and x1v9. The families q0 and x1v8 play a major role in model
  A3 and thus could be essential for fast rotating bars.
  Nevertheless, the geometry of a boxy feature becomes
  evident mainly in
  weighted profiles, where orbits from one or more families are 
  considered. Inner boxiness is associated mainly with
  the x1v1 family and, in model B, with family z3.1s.
\item Orbits of the families related to the 3:1 resonance (t1 and t2) may
  contribute in some cases to the boxiness of the profiles. The t1 family
  can do this by supporting motion parallel to the minor axis of the bar in
  intermediate 
  distances between the galaxy center and corotation, and the t2
  family by enhancing the sides of the boxes parallel to the major axis.
\item The consideration of several families of orbits for building a profile 
  may lead to boxy features close to the end of bars, with
  $p_{max}/p_{min}$ ratios different from the corresponding ratios of
  individual orbits or families of orbits.
\item An essential conclusion of our investigation is that outer boxiness is
  favoured by fast bar pattern speeds, while in the slow-rotating model
  the bar is surrounded by almost circular orbits. These are the two extremes
  of an orbital behaviour that changes as the pattern speed varies from one
  model to the other. The near-circular orbits could be building
  blocks for inner rings.
\item The fast rotating bar has a length 0.95 of its corotation radius, while
  the slow one only 0.68. This indicates that boxy bars end close to their
  corotation, while the end in slow rotating bars might be associated with
  $n:1$ resonances of lower $n$ values.
\item The rectangular-like orbits in models with faster bars are more
  elongated than the corresponding orbits in models with slower bars. When
  modeling individual barred galaxies, for which we know a priori that in
  general $1<R_c/a_o<1.4$, the elongation of the orbits can be used to give
  estimates of, or at least set limits to, the bar pattern speed, without the
  use of kinematics, which is not always available.
\item The second most efficient way of stretching rectangular-like
  orbits is to increase
  the mass of the bar. This mechanism, however, is restricted by the fact that
  in very strong bars  the
  families of periodic orbits which support the boxy face-on profiles
  are unstable over large energy intervals.
\item Outer boxy, `bow tie' morphology is possible in some models by combining
  orbits of several families. A necessary condition for this is to have 4:1
  type orbits with loops close to the main bar. Inner `bow tie' morphology can
  be due to the presence of x1v1 orbits (Fig.~\ref{A1xy}c, Fig.~\ref{A3xy}c,
  Fig.~\ref{Hxy}b).
 
\end{enumerate}

\section*{Acknowledgments}
We acknowledge fruitful discussions and very useful comments by G.~Contopoulos
and A.~Bosma. We thank the anonymous referee for valuable remarks, which
improved the paper.
This work has been supported by E$\Pi$ET II and K$\Pi\Sigma$
1994-1999; and by the Research Committee of the Academy of Athens. ChS and PAP
thank the Laboratoire d'Astrophysique de Marseille, for an invitation during
which, essential parts of this work have been completed.  ChS was partially
supported by the ``Karatheodory'' post-doctoral fellowship No 2794 of the
University of Patras.  All image processing work has been done with ESO-MIDAS.
DSS was produced at STSI under U.S. Government grant NAG W-2166. The images
of these surveys are based on photographic data obtained using the Oschin
Schmidt Telescope on Palomar Mountain and the UK Schmidt Telescope.

\bsp

\label{lastpage}

\end{document}